\newcommand{\COBOLD}{{\sf CO5BOLD}}

\newcommand{\COROT}{{ CoRoT}}
\newcommand{\MOST}{{\sc Most}}
\newcommand{\KEPLER}{{\it Kepler}}
\newcommand{\LHD}{{\sf LHD}}

\newcommand{\pun}[1]{\mbox{\rm\,#1}}
\newcommand{\mlp}{\ensuremath{\alpha_{\mathrm{MLT}}}}
\newcommand{\Teff}{\ensuremath{T_{\mathrm{eff}}}}
\newcommand{\logg}{\ensuremath{\log g}}

\newcommand{\beq}{\begin{equation}}
\newcommand{\eeq}{\end{equation}}

\newcommand{\tauross}{\ensuremath{\tau_\mathrm{Ross}}}
\newcommand{\senv}{\ensuremath{\mathrm{s}_{\mathrm{env}}}}
\newcommand{\fturb}{\ensuremath{f_\mathrm{turb}}}
\newcommand{\Pturb}{\ensuremath{P_\mathrm{turb}}}
\newcommand{\vconv}{\ensuremath{v_\mathrm{c}}}

\newcommand{\moh}{\ensuremath{\left[\mathrm{M/H}\right]}}

\newcommand\araa{{ARA\&A\,}}%
\newcommand\apj{{ApJ\,}}%
\newcommand\apjl{{ApJ\,}}%
\newcommand\apjs{{ApJS\,}}%
\newcommand\apss{{Ap\&SS\,}}%
\newcommand\aap{{A\&A\,}}%
\newcommand\ssr{{Space Science Rev.\,}}%
\newcommand\jcp{{J.Comp.Phys.\,}}%
\newcommand\memsai{{MmSAI\,}}%

\documentclass[graybox, natbib, footinfo]{svmult}

\usepackage{mathptmx}       
\usepackage{helvet}         
\usepackage{courier}        
\usepackage{type1cm}        
\usepackage{makeidx}         
\usepackage{graphicx}        
\usepackage{multicol}        
\usepackage[bottom]{footmisc}
\usepackage{url}

\makeindex             

\begin{document}

\title*{3D Model Atmospheres of Red Giant Stars}

\author{Hans-G\"unter Ludwig and Matthias Steffen}
\institute{H.-G. Ludwig \at Zentrum f\"ur Astronomie der Universit\"at
  Heidelberg, Landessternwarte, K\"onigstuhl 12, D-69117 Heidelberg, Germany,
  \email{H.Ludwig@lsw.uni-heidelberg.de}
\and
M. Steffen \at Astrophysikalisches Institut Potsdam,
An der Sternwarte 16,
D-14482~Potsdam,
Germany,
\email{msteffen@aip.de}}

\maketitle

\abstract{We provide a brief overview of the modelling of the atmospheres of
  red giant stars with the 3D radiation-hydrodynamics code \COBOLD. We
  emphasize aspects where 3D modelling provides additional insight beyond
  standard hydrostatic 1D models, and comment on present modelling
  challenges.}

\section{Introduction}

Modelling of three-dimensional (3D) atmospheres of cool stars is an active field of
development \citep[e.g.][]{Nagendra+al09}, and particularly 3D models of
atmospheres of red giant (RG) stars are just on the verge of becoming
available for application to astrophysical problems. In an early
application, \cite{Kucinskas+al05} used a 3D RG model to estimate color
corrections due to thermal inhomogeneities; \cite{Collet+al07} considered a
set of eight giant models to investigate the impact on line formation and
abundances. More recently, \cite{Freytag+Hoefner08} developed model
atmospheres of AGB stars and their winds, \cite{Dupret+al09}
derived the energy input to solar-like oscillations in giants from 3D models,
\cite{Ramirez+al10} studied convective line-shifts in the metal-poor RG
HD\,122563 and compared them to a 3D model, \cite{Chiavassa+al11} applied
global 3D models to assess effects of photometric and related astrometric
variability, and \cite{Pasquini+al11} took recourse to 3D dwarf and RG models
to correct for convective blueshifts in high-precision, spectroscopic radial
velocity measurements. While fairly exhaustive, the list of examples is still
quite short, but illustrates already the variety of possible
applications of 3D RG models. At the moment efforts are under way to cover the
Hertzsprung-Russell diagram with 3D model atmospheres including stars in the
red-giant branch \citep{Ludwig+al09b,Trampedach+Stein11}.

In the following we are going to focus on aspects related to abundances from
3D models, and the theoretical calibration of the mixing-length
parameter~\mlp\ from 3D model atmospheres. Here, we are not so much presenting
results as rather pointing out problems which are still lingering. 
We finally add
some comments about predictions of the photometric micro-variability which are
of interest in the context of high-precision photometry missions like \COROT.

\section{Our model atmosphere codes}
\subsection{3D model atmospheres -- \COBOLD}

Our 3D model atmospheres were calculated with the radiation-hydrodynamics code
\COBOLD\ \citep{Freytag+al02,Wedemeyer+al04,Freytag+al11}.  The code solves
the time-dependent equations of compressible hydrodynamics coupled to
radiative transfer in a constant gravity field in a Cartesian computational
domain which is representative of a volume located at the stellar surface.
The equation of state takes into consideration the ionization of hydrogen and
helium, as well as the formation of H$_2$ molecules according to
Saha-Boltzmann statistics. Relevant thermodynamic quantities -- in particular
gas pressure and temperature -- are tabulated as a function of gas density and
internal energy.  The multi-group opacities used by \COBOLD\ are based on
monochromatic opacities stemming from the MARCS stellar atmosphere package
\citep{Gustafsson+al08} provided as function of gas pressure and temperature
with high wavelength resolution. The opacities have been calculated 
assuming solar
elemental abundances according to \cite{Grevesse+Sauval98}, with the
exception of CNO for which values close to the recommendation of
\cite{Asplund05} are adopted (specifically, A(C)=8.41, A(N)=7.80,
A(O)=8.67). The metal abundances were scaled according to overall metallicity
of the model assuming an enhancement of the $\alpha$-elements by +0.4\,dex at
metallicities $\moh<-1$.

In our RG models we typically use a number of $140\times 140 \times 150$ to
$160\times 160\times 200$ points for the hydrodynamical grid. The decision
about the resolution primarily hinges on the effective temperature of the
model, hotter models usually require a higher resolution. The wavelength
dependence of the radiation field is represented by 5 multi-group bins in the
case of solar metallicity, and 6 bins at sub-solar metallicities, following the
procedures laid out by
\citet{Nordlund82,Ludwig92,Ludwig+al94,Voegler+al04}. For test purposes we
have calculated a few models with more bins. Since it is of relevance for the
discussion later, we emphasize that all opacity sources -- including scattering
opacities -- are treated as true absorption. The sorting into wavelength groups is done
applying thresholds in logarithmic Rosseland optical depth $\{ +\infty, 0.0,
-1.5, $ $-3.0, -4.5, -\infty\}$ for the 5-bin, and $\{+\infty, 0.1, 0.0,$
$-1.0, -2.0, -3.0, -\infty\}$ for the 6-bin schemes. In all but one bin a
switching between Rosseland and Planck averages is performed at a
band-averaged Rosseland optical depth of 0.35; in the bin gathering the largest
line opacities, the Rosseland mean opacity is used throughout. The decisions
about number of bins, and sorting thresholds are motivated by comparing
radiative fluxes and heating rates obtained by the binned opacities in
comparison to the case of high wavelength resolution.

\subsection{1D stellar atmospheres -- \LHD}

Due to still present limitations in the realism (e.g. by the limited
wavelength resolution) of 3D model atmospheres it is often advantageous to
work differentially, and express 3D effects relative to a 1D comparison
structure. To this end we developed a 1D stellar atmosphere code called
\LHD\ which employs the same opacities and equation-of-state as the 3D code
\COBOLD. The convective energy transport is modelled in the framework of
mixing-length theory as described in \cite{Mihalas78}. The resulting 1D
stratifications are in hydrostatic and radiative-convective equilibrium. See
\cite{Caffau+al07} for more details on our approach of deriving abundance
corrections.

\section{RG abundances and the issue of scattering}

\cite{Collet+al07} presented 3D-1D abundance corrections for RG models at
effective temperatures of around 5000\,K, $\logg=2.2$, and metallicities
ranging from solar to $\moh=-3$, using the 3D code of Nordlund \&\ Stein
\citep{Stein+Nordlund98}. Two similar studies were presented by
\cite{Dobrovolskas+al10} and \cite{Ivanauskas+al10} who used \COBOLD\ and
\LHD\ models at about $\Teff=5000\,\mbox{K}$, $\logg=2.5$, with metallicities
down to $\moh=-3$ to derive 3D-1D abundance corrections. 
While the two later studies confirm the results of Collet and collaborators,
showing that generally the magnitude
of 3D-1D abundance corrections becomes larger towards lower metallicity, the
quantitative agreement is not satisfactory. The \COBOLD-based abundance
corrections are usually noticeably smaller in magnitude, in particular at the
lowest metallicities. Obviously, this is an unfortunate situation, and one
would like to see a higher degree of consistency among results from different
3D codes.

In a recent paper, \cite{Collet+al11} suggested the treatment of scattering in
the simulations as the reason for the discrepant abundance corrections for RGs 
at low metallicity. The main scattering process is Rayleigh scattering by 
neutral hydrogen. This is perhaps the simplest case of scattering and can be 
modelled as coherent isotropic scattering in the continuum. Collet and 
collaborators implemented a proper treatment of this kind of scattering in 3D.
They also put
forward an approximate treatment of scattering by simply leaving out scattering
contributions in the binned opacities in the optically thin regions. They showed
that this approximate treatment provides results in close agreement with the
exact treatment. They further performed a comparison with the case where
scattering is treated as true absorption -- as is the case in the
\COBOLD\ models. Their models of 2007 used the approximate treatment of
scattering. The models show a sensitive dependence of the resulting
temperature stratification on the treatment of scattering. In their $\moh=-3$
RG model, the difference amounts to 600\,K at optical depth $\log\,\tauross=-4$
in the sense that a proper treatment of scattering leads to cooler structures
in comparison to treating scattering opacities as true absorption.

\begin{figure}
\centering
\includegraphics[width=\textwidth]{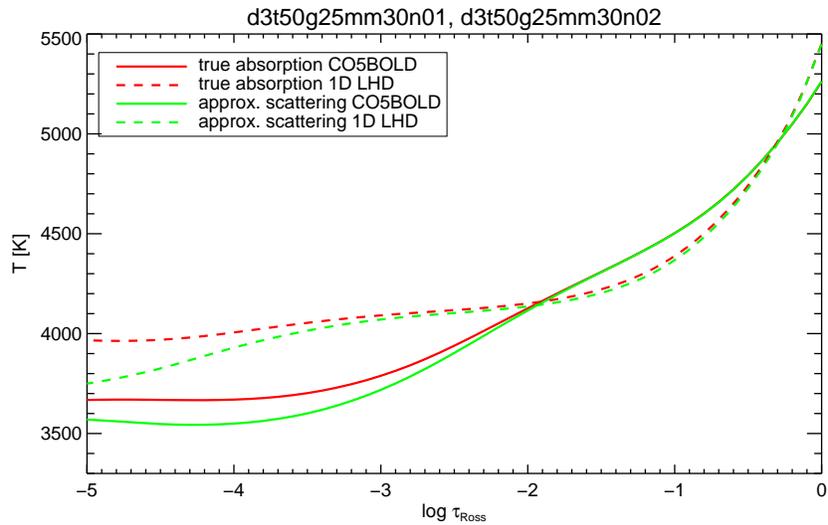}

\vspace{-\baselineskip}

\caption{Comparison of the mean temperature structures of two different 3D
  \COBOLD\ hydrodynamical model atmospheres (solid) and associated 1D LHD models
  (dashed), computed with a different treatment of radiative transfer. In the
  first case (dark [red] curves), the continuum scattering opacity is treated
  as true absorption opacity, while in the second case (light [green] curves),
  the continuum scattering opacity is ignored in the optically thin
  layers. For the 3D models, averaging was performed on surfaces of constant
  Rosseland optical depth and over 70 equidistant snapshots covering a total
  of $140\,000$~s.}
\label{f:1}
\end{figure}

\begin{figure}
\centering
\includegraphics[width=\textwidth]{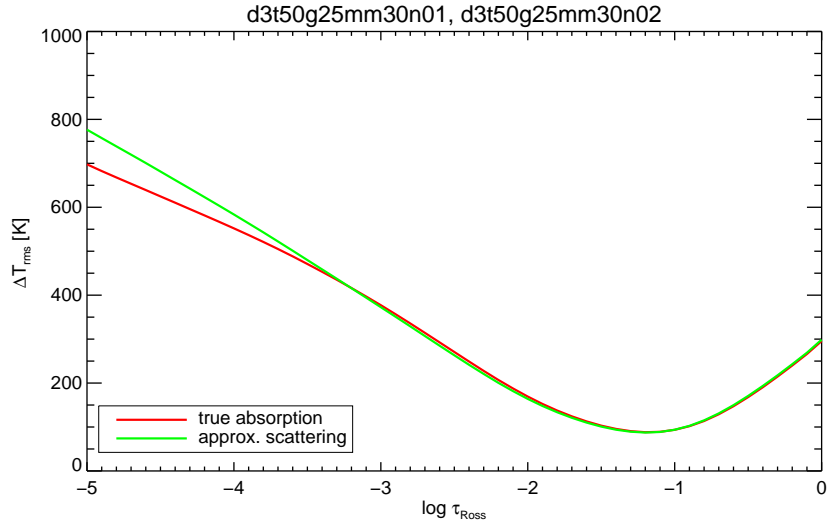}

\vspace{-\baselineskip}

\caption{Total rms temperature fluctuation $\Delta T_{\rm rms, tot}$ as a
  function of Rosseland optical depth for the two 3D models shown in 
  Fig.~\ref{f:1}, computed as 
  $\Delta T_{\rm rms, tot}=\sqrt{ \langle T^2\rangle_{x,y,t}
    - \langle T\rangle_{x,y,t}^2 }$, where $\langle.\rangle_{x,y,t}$ denotes
  horizontal averaging over surfaces of constant Rosseland optical depth and
  over time. We have verified that the amplitude of the total temperature
  fluctuation is completely dominated by the spatial temperature variations: 
  $\Delta T_{\rm rms,tot} \approx \Delta T_{\rm rms, xy} = 
  \left\langle\sqrt{ \langle T^2\rangle_{x,y} - \langle T\rangle_{x,y}^2 }\,
  \right\rangle_t$.}
\label{f:2}
\end{figure}

It appears plausible that the differing 3D-1D abundance corrections are a
consequence of the different thermal structures resulting from the different
treatment of scattering in the \COBOLD\ and Nordlund-Stein class of models. To
test this idea, we calculated a RG model with the same atmospheric parameters
as before but with the approximate treatment of scattering as suggested by
\cite{Collet+al11}. Figures~\ref{f:1} and~\ref{f:2} illustrate the
outcome. The most striking aspect is that our models show a very much reduced
sensitivity to the treatment of scattering in comparison to the models of
Collet and co-workers. The approximate treatment of scattering leads to a
structure which is only 120\,K cooler at $\log\,\tauross=-4$, in comparison to
$\approx 600\,\mbox{K}$ found by \cite{Collet+al11}. 
This also carries over to the
temperature fluctuations which are little affected by the treatment of
scattering (see Fig.~\ref{f:2}). We already emphasized the importance of a
differential approach, and Fig.~\ref{f:1} also shows the effects on the 
associated 1D \LHD\ models. Temperature \emph{differences} between the 
1D and 3D models at given
optical depth are changing even less. While we did not perform spectrum
synthesis calculations yet to derive new abundance corrections, we consider
it as unlikely that the modest changes in the thermal structure can change our
abundance corrections so much that they become consistent with the values of
\cite{Collet+al07}.

The situation remains puzzling. The very different sensitivity to the
treatment of scattering is difficult to explain. We only can hint at the
differences in the calculation of the band-averaged opacities in the various 
codes: Collet and collaborators use intensity-averaged opacities in the 
optically thin regions, while we use Planck-averages -- except for the band 
collecting the strongest lines where a Rosseland average is used throughout. 
We speculate that these choices, together with the definition of the opacity 
bins, may have a significant influence on the resulting thermal structures 
and their sensitivity to the treatment of scattering.

\section{The calibration of \mlp\ and turbulent pressure}

It is well known from the theory of stellar structure that convection is
generally an efficient means of transporting energy, and that it establishes a
thermal structure close to adiabatic. Only in the vicinity of the boundaries of
convective regions noticeable deviations from adiabaticity occur.  In
convective envelopes of late-type stars the upper boundary of the convective
envelope -- usually located close to or even in the optically thin layers --
constitutes the bottle-neck for the energy transport through the stellar
envelope assigning a special role to it. Despite its small geometrical extent
and low mass, it largely determines the properties of the convective
envelope as a whole. It is the value of the entropy of the adiabatically
stratified bulk of the convective~\senv\ which is most important from the
point of view of stellar structure since it influences the resulting radius
and effective temperature of a stellar model. \senv\ is controlled by the
efficiency of convective and radiative energy transport in the thin,
superadiabatically stratified surface layers. 3D model atmospheres can be
applied to model this region, and allow to quantify the mutual efficiency of the
convective and radiative energy transport, and to predict \senv.  Comparing the
model predictions to standard 1D models based on mixing-length theory (MLT) the
value of \senv\ can be translated into a corresponding mixing-length
parameter~\mlp\ \citep{Trampedach+al99,Ludwig+al99,Ludwig+al08}.

In stellar evolution calculations the free mixing-length parameter is usually
calibrated against the Sun. However, it is unclear whether mixing-length
theory provides a suitable scaling of the convective efficiency at constant
\mlp\ across the Hertzsprung-Russell diagram. The depth of the surface
convective envelope and the related \mlp\ can be constraint by
asteroseismology. However, degeneracies with other parameters often make it
difficult to obtain a unique solution \citep[e.g.][]{Goupil+al11}. Hence, it
would be useful to have an independent estimate available which 3D models can
provide in principle.

\begin{figure}
\begin{center}
\includegraphics[width=0.6\textwidth]{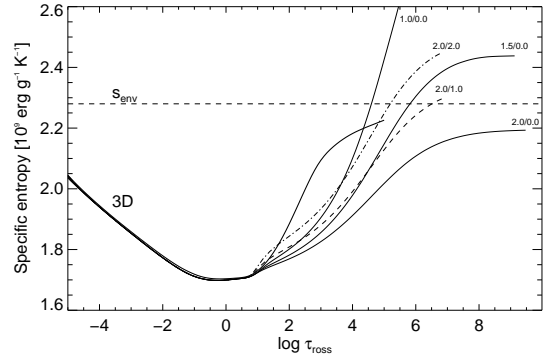}
\end{center}
\vspace{-\baselineskip}
  \caption{Entropy-optical depth profiles (horizontal and temporal average) of
    a 3D red giant model (thick solid line) in comparison to 1D stellar
    atmosphere models of different \mlp\ leaving out (thin solid lines) or
    including (dashed and dashed-dotted lines) turbulent pressure. The lines
    are labeled with values \mbox{\mlp/\fturb}
    (details see text). The horizontal dashed line indicates the value of
    \senv\ as predicted by the 3D model.}
\label{f:pturb}
\end{figure}

In main-sequence models turbulent pressure plays generally only a minor role
but becomes relatively more important towards lower gravities -- and causes
trouble when one is interested in a well-defined calibration of the
mixing-length parameter. Figure~\ref{f:pturb} shows the average temperature
profile of a 3D red giant model ($\Teff\approx 3600\,\mathrm{K}$,
\mbox{\logg=1.0}, \mbox{\moh=0.0}) in comparison to standard 1D model
atmospheres of the same atmospheric parameters.  While the turbulent
pressure~\Pturb\ is naturally included in the 3D models, it is modelled in a
ad-hoc fashion in 1D models, assuming a parameterisation
$\Pturb=\fturb\rho\vconv^2$, where \fturb\ is a free parameter of order unity,
$\rho$ the mass density and \vconv\ the convective velocity according MLT.

Figure~\ref{f:pturb} shows that it is essentially impossible to reproduce the
mean thermal profile of the 3D model with any of the 1D models -- irrespective
of the choice of \mlp\ and \fturb. The failure is related to the local nature
of MLT confining the action of the turbulent pressure gradients strictly to
the convectively unstable regions. While formally one can still match
\senv\ of the 3D model by a 1D profile with suitably chosen \mlp\ and/or
\fturb\ such a match becomes physically little motivated, and is unlikely to
provide a robust scaling with changing atmospheric parameters. An improved 1D
convection description including non-local effects like overshooting is
clearly desirable to handle this situation.  Empirical calibrations of
\mlp\ using giants are likely to suffer from ambiguities related to the way
turbulent pressure is treated in the 1D models. One may take the result as
an indication that taking recourse to 1D models is not warranted, and one may
give up the benefits of a differential approach by relating 3D to 1D
structures. Alternatively, one may take the absolute entropy of the convective
envelope (perhaps translated to equivalent pressure-temperature pairs) as
predicted by the 3D model as constraint to be matched in 1D stellar structure
models.

\section{Granulation-related photometric micro-variability}

High-precision photometry of satellite missions (foremost \MOST, \COROT,
\KEPLER) allow the detection of stellar variability associated with the random
changes of the granulation pattern on the surfaces of late-tape stars -- by
asteroseismologists usually referred to as ``granular background noise''. 3D
model atmospheres represent the granulation pattern in detail and allow to
predict the power spectrum of the variability signal
\citep{Trampedach+al98,Svensson+Ludwig05,Ludwig06}. Despite this possibility,
no comprehensive theoretical study has been conducted so far. One of the
reasons is that long time series need to be calculated to collect sufficient
statistics, which is computationally demanding. The F-dwarf HD\,49933 -- a
prominent \COROT-target -- is an exception for which \cite{Ludwig+al09}
performed an analysis. However, the growing body of observational data in
particular for giant stars should motivate further efforts in this direction.
Recently, \cite{Kjeldsen+Bedding11} suggested a new scaling relation for the
amplitude of solar-like oscillations, and also discuss the scaling of the
granulation background signal. It would be interesting to see whether 3D model
atmospheres can lend further support to the suggested relations.

\begin{figure}
\begin{center}
\includegraphics[width=0.5\textwidth, angle=90]{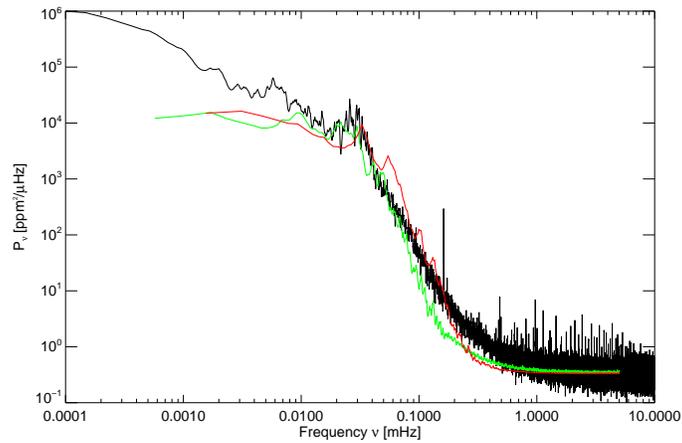}
\end{center}
\vspace{-0.5\baselineskip}
  \caption{Comparison of power spectra of photometric variability. Black line: HD\,181907 as observed by
    \COROT. The predicted power spectra from two 3D
    models are depicted with light [green ] ($\Teff=4500\pun{K}$) and dark
    grey [red] ($\Teff=5000\pun{K}$) lines. Further details see text.}
\label{f:mvar}
\end{figure}

To illustrate the feasibility, we show in Fig.~\ref{f:mvar} a rough comparison
of the photometric variability between the RG HD\,181907 (HR\,7349) and
predictions from two 3D models. The plot focuses on the frequency region where
the granulation-related signal is expected. \COROT\ acquired a high-quality
time series for HD\,181907; \cite{Carrier+al10} give atmospheric parameters
$4780\pm 80$/$2.78\pm 0.16$/$-0.08\pm 0.10$ (\Teff/\logg/\moh). The two 3D
models have atmospheric parameters 4500/2.5/0.0 and 5000/2.5/0.0, bracketing
the star in effective temperature, as well as having comparable surface
gravity and metallicity. Although no dedicated modelling was performed, the
spectra appear quite similar.

\section{Concluding remarks}

3D model atmospheres of cool stars, including red giants,
have reached a level of realism which allows a direct
confrontation with observations. In some areas they allow to make predictions
beyond the capabilities of classical 1D models. However, as we have seen there
still exist modelling challenges, and last but not least quite some work is
still necessary to fully exploit the potential of such models.


\begin{thebibliography}{36}
\expandafter\ifx\csname natexlab\endcsname\relax\def\natexlab#1{#1}\fi

\bibitem[{{Asplund}(2005)}]{Asplund05}
{Asplund}, M. 2005, \araa, 43, 481

\bibitem[{{Caffau} {et~al.}(2007){Caffau}, {Faraggiana}, {Bonifacio}, {Ludwig},
  \& {Steffen}}]{Caffau+al07}
{Caffau}, E., {Faraggiana}, R., {Bonifacio}, P., {Ludwig}, H.-G., \& {Steffen},
  M. 2007, \aap, 470, 699

\bibitem[{{Carrier} {et~al.}(2010){Carrier}, {Morel}, {Miglio},
  {Montalb{\'a}n}, {Auvergne}, {Baglin}, {Baudin}, {Barban}, {Catala},
  {D'Antona}, {De Ridder}, {Eggenberger}, {Hatzes}, {Hekker}, {Kallinger},
  {Michel}, {Noels}, {Poretti}, {Rainer}, {Samadi}, {Ventura}, \&
  {Weiss}}]{Carrier+al10}
{Carrier}, F., {Morel}, T., {Miglio}, A., {et~al.} 2010, \apss, 328, 83

\bibitem[{{Chiavassa} {et~al.}(2011){Chiavassa}, {Pasquato}, {Jorissen},
  {Sacuto}, {Babusiaux}, {Freytag}, {Ludwig}, {Cruzal{\`e}bes}, {Rabbia},
  {Spang}, \& {Chesneau}}]{Chiavassa+al11}
{Chiavassa}, A., {Pasquato}, E., {Jorissen}, A., {et~al.} 2011, \aap, 528, A120

\bibitem[{{Collet} {et~al.}(2007){Collet}, {Asplund}, \&
  {Trampedach}}]{Collet+al07}
{Collet}, R., {Asplund}, M., \& {Trampedach}, R. 2007, \aap, 469, 687

\bibitem[{{Collet} {et~al.}(2011){Collet}, {Hayek}, {Asplund}, {Nordlund},
  {Trampedach}, \& {Gudiksen}}]{Collet+al11}
{Collet}, R., {Hayek}, W., {Asplund}, M., {et~al.} 2011, \aap, 528, A32

\bibitem[{{Dobrovolskas} {et~al.}(2010){Dobrovolskas}, {Ku{\v c}inskas},
  {Ludwig}, {Caffau}, {Klevas}, \& {Prakapavi{\v c}ius}}]{Dobrovolskas+al10}
{Dobrovolskas}, V., {Ku{\v c}inskas}, A., {Ludwig}, H.-G., {et~al.} 2010, ArXiv
  e-prints

\bibitem[{{Dupret} {et~al.}(2009){Dupret}, {Belkacem}, {Samadi}, {Montalban},
  {Moreira}, {Miglio}, {Godart}, {Ventura}, {Ludwig}, {Grigahc{\`e}ne},
  {Goupil}, {Noels}, \& {Caffau}}]{Dupret+al09}
{Dupret}, M., {Belkacem}, K., {Samadi}, R., {et~al.} 2009, \aap, 506, 57

\bibitem[{{Freytag} \& {H{\"o}fner}(2008)}]{Freytag+Hoefner08}
{Freytag}, B. \& {H{\"o}fner}, S. 2008, \aap, 483, 571

\bibitem[{{Freytag} {et~al.}(2002){Freytag}, {Steffen}, \&
  {Dorch}}]{Freytag+al02}
{Freytag}, B., {Steffen}, M., \& {Dorch}, B. 2002, Astronomische Nachrichten,
  323, 213

\bibitem[{{Freytag} {et~al.}(2011){Freytag}, {Steffen}, {Ludwig},
  {Wedemeyer-B{\"o}hm}, {Schaffenberger}, \& {Steiner}}]{Freytag+al11}
{Freytag}, B., {Steffen}, M., {Ludwig}, H.-G., {et~al.} 2011, \jcp, submitted

\bibitem[{{Goupil} {et~al.}(2011){Goupil}, {Lebreton}, {Marques}, {Deheuvels},
  {Benomar}, \& {Provost}}]{Goupil+al11}
{Goupil}, M.~J., {Lebreton}, Y., {Marques}, J.~P., {et~al.} 2011, Journal of
  Physics Conference Series, 271, 012032

\bibitem[{{Grevesse} \& {Sauval}(1998)}]{Grevesse+Sauval98}
{Grevesse}, N. \& {Sauval}, A.~J. 1998, \ssr, 85, 161

\bibitem[{{Gustafsson} {et~al.}(2008){Gustafsson}, {Edvardsson}, {Eriksson},
  {J{\o}rgensen}, {Nordlund}, \& {Plez}}]{Gustafsson+al08}
{Gustafsson}, B., {Edvardsson}, B., {Eriksson}, K., {et~al.} 2008, \aap, 486,
  951

\bibitem[{{Ivanauskas} {et~al.}(2010){Ivanauskas}, {Kucinskas}, {Ludwig}, \&
  {Caffau}}]{Ivanauskas+al10}
{Ivanauskas}, A., {Kucinskas}, A., {Ludwig}, H.~G., \& {Caffau}, E. 2010, in
  Nuclei in the Cosmos.

\bibitem[{{Kjeldsen} \& {Bedding}(2011)}]{Kjeldsen+Bedding11}
{Kjeldsen}, H. \& {Bedding}, T.~R. 2011, \aap, 529, L8

\bibitem[{{Ku{\v c}inskas} {et~al.}(2005){Ku{\v c}inskas}, {Hauschildt},
  {Ludwig}, {Brott}, {Vansevi{\v c}ius}, {Lindegren}, {Tanab{\'e}}, \&
  {Allard}}]{Kucinskas+al05}
{Ku{\v c}inskas}, A., {Hauschildt}, P.~H., {Ludwig}, H.-G., {et~al.} 2005,
  \aap, 442, 281

\bibitem[{{Ludwig}(1992)}]{Ludwig92}
{Ludwig}, H.-G. 1992, PhD thesis, University of Kiel

\bibitem[{{Ludwig}(2006)}]{Ludwig06}
{Ludwig}, H.-G. 2006, \aap, 445, 661

\bibitem[{{Ludwig} {et~al.}(2008){Ludwig}, {Caffau}, \& {Ku{\v
  c}inskas}}]{Ludwig+al08}
{Ludwig}, H.-G., {Caffau}, E., \& {Ku{\v c}inskas}, A. 2008, in IAU Symposium,
  Vol. 252, IAU Symposium, ed. {L.~Deng \& K.~L.~Chan}, 75--81

\bibitem[{{Ludwig} {et~al.}(2009{\natexlab{a}}){Ludwig}, {Caffau}, {Steffen},
  {Freytag}, {Bonifacio}, \& {Ku{\v c}inskas}}]{Ludwig+al09b}
{Ludwig}, H.-G., {Caffau}, E., {Steffen}, M., {et~al.} 2009{\natexlab{a}},
  \memsai, 80, 711

\bibitem[{{Ludwig} {et~al.}(1999){Ludwig}, {Freytag}, \&
  {Steffen}}]{Ludwig+al99}
{Ludwig}, H.-G., {Freytag}, B., \& {Steffen}, M. 1999, \aap, 346, 111

\bibitem[{{Ludwig} {et~al.}(1994){Ludwig}, {Jordan}, \&
  {Steffen}}]{Ludwig+al94}
{Ludwig}, H.-G., {Jordan}, S., \& {Steffen}, M. 1994, \aap, 284, 105

\bibitem[{{Ludwig} {et~al.}(2009{\natexlab{b}}){Ludwig}, {Samadi}, {Steffen},
  {Appourchaux}, {Baudin}, {Belkacem}, {Boumier}, {Goupil}, \&
  {Michel}}]{Ludwig+al09}
{Ludwig}, H.-G., {Samadi}, R., {Steffen}, M., {et~al.} 2009{\natexlab{b}},
  \aap, 506, 167

\bibitem[{{Mihalas}(1978)}]{Mihalas78}
{Mihalas}, D. 1978, {Stellar atmospheres, 2nd edition}

\bibitem[{{Nagendra} {et~al.}(2009){Nagendra}, {Bonifacio}, \&
  {Ludwig}}]{Nagendra+al09}
{Nagendra}, K.~N., {Bonifacio}, P., \& {Ludwig}, H.-G. 2009, \memsai, 80, 601

\bibitem[{{Nordlund}(1982)}]{Nordlund82}
{Nordlund}, A. 1982, \aap, 107, 1

\bibitem[{{Pasquini} {et~al.}(2011){Pasquini}, {Melo}, {Chavero}, {Dravins},
  {Ludwig}, {Bonifacio}, \& {de La Reza}}]{Pasquini+al11}
{Pasquini}, L., {Melo}, C., {Chavero}, C., {et~al.} 2011, \aap, 526, A127

\bibitem[{{Ram{\'{\i}}rez} {et~al.}(2010){Ram{\'{\i}}rez}, {Collet}, {Lambert},
  {Allende Prieto}, \& {Asplund}}]{Ramirez+al10}
{Ram{\'{\i}}rez}, I., {Collet}, R., {Lambert}, D.~L., {Allende Prieto}, C., \&
  {Asplund}, M. 2010, \apjl, 725, L223

\bibitem[{{Stein} \& {Nordlund}(1998)}]{Stein+Nordlund98}
{Stein}, R.~F. \& {Nordlund}, A. 1998, \apj, 499, 914

\bibitem[{{Svensson} \& {Ludwig}(2005)}]{Svensson+Ludwig05}
{Svensson}, F. \& {Ludwig}, H.-G. 2005, in ESA Special Publication, Vol. 560,
  13th Cambridge Workshop on Cool Stars, Stellar Systems and the Sun, ed.
  {F.~Favata, G.~A.~J.~Hussain, \& B.~Battrick}, 979

\bibitem[{{Trampedach} {et~al.}(1998){Trampedach}, {Christensen-Dalsgaard},
  {Nordlund}, \& {Stein}}]{Trampedach+al98}
{Trampedach}, R., {Christensen-Dalsgaard}, J., {Nordlund}, A., \& {Stein},
  R.~F. 1998, in The First MONS Workshop: Science with a Small Space Telescope,
  ed. {H.~Kjeldsen \& T.~R.~Bedding}, 59

\bibitem[{{Trampedach} \& {Stein}(2011)}]{Trampedach+Stein11}
{Trampedach}, R. \& {Stein}, R.~F. 2011, \apj, 731, 78

\bibitem[{{Trampedach} {et~al.}(1999){Trampedach}, {Stein},
  {Christensen-Dalsgaard}, \& {Nordlund}}]{Trampedach+al99}
{Trampedach}, R., {Stein}, R.~F., {Christensen-Dalsgaard}, J., \& {Nordlund},
  {\AA}. 1999, in Astronomical Society of the Pacific Conference Series, Vol.
  173, Stellar Structure: Theory and Test of Connective Energy Transport, ed.
  {A.~Gimenez, E.~F.~Guinan, \& B.~Montesinos}, 233

\bibitem[{{V{\"o}gler} {et~al.}(2004){V{\"o}gler}, {Bruls}, \&
  {Sch{\"u}ssler}}]{Voegler+al04}
{V{\"o}gler}, A., {Bruls}, J.~H.~M.~J., \& {Sch{\"u}ssler}, M. 2004, \aap, 421,
  741

\bibitem[{{Wedemeyer} {et~al.}(2004){Wedemeyer}, {Freytag}, {Steffen},
  {Ludwig}, \& {Holweger}}]{Wedemeyer+al04}
{Wedemeyer}, S., {Freytag}, B., {Steffen}, M., {Ludwig}, H.-G., \& {Holweger},
  H. 2004, \aap, 414, 1121

\end{thebibliography}
\end{document}